\documentclass{ws-procs9x6}
\usepackage{amsmath,amssymb}

\newcommand{\mathbbm}[1]{\Bbb{#1}}

\newcommand{\bigintlim}{\int}
\newcommand{\bignone}{\,}

\newcommand{\mathd}{\mathrm{d}}
\newcommand{\mathe}{\mathrm{e}}

\newcommand{\tmmathbf}[1]{\ensuremath{\boldsymbol{#1}}}
\newcommand{\tmop}[1]{\ensuremath{\text{#1}}}
\newcommand{\tmtextbf}[1]{{\bfseries{#1}}}
\newcommand{\tmtextit}[1]{{\itshape{#1}}}

\newcommand{\TeXmacs}{T\kern-.1667em\lower.5ex\hbox{E}\kern-.125emX\kern-.1em\lower.5ex\hbox{\textsc{m\kern-.05ema\kern-.125emc\kern-.05ems}}}
\newenvironment{itemizedot}{\begin{itemize} \renewcommand{\labelitemi}{$\bullet$}}{\end{itemize}}

\begin{document}

\title{DESCRIPTION OF DECOHERENCE BY MEANS OF TRANSLATION-COVARIANT MASTER EQUATIONS AND L\'EVY PROCESSES}%
\author{BASSANO VACCHINI}%
\address{Dipartimento di Fisica
dell'Universit\`a di Milano and INFN,
Sezione di Milano
\\
Via Celoria 16, 20133, Milan, Italy}

\begin{abstract}
Translation-covariant Markovian master equations used in the
description of decoherence and dissipation are considered in
the general framework of Holevo's results on the
characterization of generators of covariant quantum dynamical
semigroups. A general connection between the characteristic
function of classical L\'evy processes and loss of coherence of
the statistical operator describing the center of mass degrees of
freedom of a quantum system interacting through momentum
transfer events with an environment is established. The
relationship with both microphysical models and experimental
realizations is considered, focusing in particular on recent
interferometric experiments exploring the boundaries between
classical and quantum world.
\end{abstract}


\bodymatter

\section{Introduction}

A natural standpoint about quantum mechanics, which is however not the one
usually considered in textbooks written for physics students, is to look at it
as a new probability theory, different and reacher than the classical one
{\cite{Streater2000a}}. This point of view becomes mandatory or at least very
fruitful if one is faced with more advanced research topics, such as the
description of open quantum systems or quantum information and communication
theory (for a general reference see {\cite{Holevo2001a,Alicki2001a}}). In
these fields tools and concepts obtained relying on a probabilistic approach,
also working in direct analogy with classical probability theory, have become
of paramount importance. An example in this direction is given by quantum
dynamical semigroups, which provide the quantum generalization of classical Markov
semigroups. The subject has been the object of active research in the
mathematics, physics and chemistry community over decades by now, but it is
still of great interest. In particular covariance properties of such mappings
under translations have been considered in detail only recently. Besides a
mathematical characterization
{\cite{Holevo1993a,Holevo1993b,Holevo1995a,Holevo1996a}} also the actual
physical relevance {\cite{Vacchini2001b,Vacchini2002b,Vacchini2005a}} of such
covariant quantum dynamical semigroups has been considered.

In the present contribution we will focus mainly on the application of such
translation-covariant quantum dynamical semigroups to the study and the
description of the phenomenon called decoherence in the physics literature
{\cite{Joos2003a,Schlosshauer2007a}}. By such a term a whole variety of
situations is meant, all having in common a loss of typical quantum
interference capability, arising as a dynamical consequence of interaction of
the system of interest with some other, typically much bigger, system. The
phenomenology of decoherence is ubiquitous when considering open quantum
systems, but its actual quantitative study requires very special experimental
conditions, which can be realized e.g. in interferometric setups for massive
particles, observing loss of interference fringes as a consequence of external
disturbance, arising because the approximation of isolation of the system is
no more realistic. For a quantitative study of the phenomenon it is in fact
crucial that such decoherence effects can actually be engineered, so that
their strength is under the control of the experimenter.

The paper is organized as follows. In Sect. \ref{sec:tv} we briefly sketch the
formal expression of the generator of a translation-covariant quantum
dynamical semigroup. In Sect. \ref{sec:lp} we show how such a general
structure in a suitable limit can account for decoherence behaviors
quantitatively described by means of the characteristic function of a
classical L\'evy process. In Sect. \ref{sec:mp} we further explore how a
particular physical example of realization of such generators applies to the
description of decoherence in both position and momentum space, finally
mentioning possible extension of the formalism in Sect. \ref{sec:co}.

\section{Translation-covariant master equations\label{sec:tv}}

Provided memory effects can be neglected, quantum dynamical semigroups
{\cite{Alicki1987a,Alicki2002b}} give a general setting for the description of
the dynamics of an open quantum system {\cite{Breuer2007}}. In the physical
literature major efforts have been devoted to the derivation or
phenomenological assessment of possible generators of such quantum dynamical
semigroups, so called master equations. The typical benchmark is the Lindblad
structure of such generators, which goes back to the work of Gorini,
Kossakowski and Sudarshan {\cite{Gorini1976a}} and of Lindblad
{\cite{Lindblad1976a}}, holding true for a generator given by a bounded mapping.
Attention was later devoted to possible constraints on the structure of such
generators arising as a consequence of symmetry (see e.g.
{\cite{Vacchini2007c}} for references). In this respect the results of Holevo
for symmetry under translations are of particular importance because of the
many possible physical applications, especially in connection with typical
quantum phenomena such as decoherence.

We first consider the general expression of formal generators of
translation-covariant quantum dynamical semigroups as obtained by Holevo
{\cite{Holevo1993a,Holevo1993b,Holevo1995a,Holevo1995b}}. The covariance of
the mapping corresponds to the requirement that its action has to commute with
the unitary representation of translations on the Hilbert space of interest.
The physical system we are going to consider is the centre of mass of a
particle in free space, so that $\mathcal{H} = L^2 \left( \mathbbm{R}^3
\right)$. Let $\mathcal{L}'$ be the mapping describing the dynamics in
Heisenberg picture, thus acting on an observable $\mathsf{A}$. In order to be
covariant $\mathcal{L}'$ has to satisfy the requirement
\begin{eqnarray}
  \mathcal{L}' \left[ \mathe^{i \mathsf{\tmmathbf{A} \cdot \mathsf{P}} /
  \hbar} \mathsf{A} \mathe^{- i \mathsf{\tmmathbf{A} \cdot \mathsf{P}} /
  \hbar} \right] & = & \mathe^{i \mathsf{\tmmathbf{A} \cdot \mathsf{P}} /
  \hbar} \mathcal{L}' \left[ \mathsf{A} \right] \mathe^{- i
  \mathsf{\tmmathbf{A} \cdot \mathsf{P}} / \hbar} \hspace{2em} \forall
  \tmmathbf{A} \in \mathbb{R}^3,  \label{eq:cov}
\end{eqnarray}
where $\mathsf{P}$ denotes the momentum operator of the massive particle. The
general structure of generator complying with this requirement is given by the
formal operator expression
\begin{eqnarray}
  \mathcal{L}' \left[ \mathsf{A}] \right. & = & \frac{i}{\hbar} [ \mathsf{H}
  \left( \mathsf{P} \right), \mathsf{A}] + \mathcal{L}_G [ \mathsf{A}] +
  \mathcal{L}_P [ \mathsf{A}],  \label{eq:g+p}
\end{eqnarray}
where the symbols $G$ and $P$ denote a Gaussian and a Poisson component, the
names arising from the connection with the classical L\'evy-Khintchine
formula. One has in particular for the Gaussian component
\begin{eqnarray}
  \mathcal{L}_G [ \mathsf{A}] & = & \frac{i}{\hbar} \left[ \mathsf{Y}_0 +
  \frac{1}{2 i} \sum_{k = 1}^3 \left( \mathsf{Y}_k L_k \left( \mathsf{P}
  \right) - L^{^{\dag}}_k \left( \mathsf{P} \right) \mathsf{Y}_k \right),
  \mathsf{A} \right] \nonumber\\
  &  & + \frac{1}{\hbar} \sum_{k = 1}^3 \left[ \left( \mathsf{Y}_k + L_k
  \left( \mathsf{P} \right) \right)^{^{\dag}} \mathsf{A} \left( \mathsf{Y}_k +
  L_k \left( \mathsf{P} \right) \right) \right. \nonumber\\
  &  & \left. - \frac{1}{2} \left\{ \left( \mathsf{Y}_k + L_k \left(
  \mathsf{P} \right) \right)^{^{\dag}} \mathsf{} \left( \mathsf{Y}_k + L_k
  \left( \mathsf{P} \right) \right), \mathsf{A} \right\} \right] \nonumber
\end{eqnarray}
where $\mathsf{Y}_j = \sum_{i = 1}^3 a_{ji} \mathsf{X}_i$ with $a_{ji} \in
\mathbbm{R}$ for $j = 0, 1, 2, 3$, that is to say it is a linear combination
of the three position operators of the test particle, appearing at most
quadratically, while for the Poisson component

\begin{eqnarray}
  \mathcal{L}_P [ \mathsf{A}] & = & \int \mathd \mu \left( \tmmathbf{Q}
  \right) \sum_j \left[ L_j^{^{\dag}} \left( \mathsf{P} ; \tmmathbf{Q} \right)
  \mathe^{- i\tmmathbf{Q} \cdot \mathsf{X} / \hbar} \mathsf{A}
  \mathe^{i\tmmathbf{Q} \cdot \mathsf{X} / \hbar} L_j \left( \mathsf{P} ;
  \tmmathbf{Q} \right) \right. \nonumber\\
  &  & \left. - \frac{1}{2} \left\{ L_j^{^{\dag}} \left( \mathsf{P} ;
  \tmmathbf{Q} \right) L_j \left( \mathsf{P} ; \tmmathbf{Q} \right),
  \mathsf{A} \right\} \right] \nonumber\\
  &  & + \int \mathd \mu \left( \tmmathbf{Q} \right) \sum_j \left[ \omega_j
  \left( \tmmathbf{Q} \right) L_j^{^{\dag}} \left( \mathsf{P} ; \tmmathbf{Q}
  \right) \left( \mathe^{- i\tmmathbf{Q} \cdot \mathsf{X} / \hbar} \mathsf{A}
  \mathe^{i\tmmathbf{Q} \cdot \mathsf{X} / \hbar} - \mathsf{A} \right) \right.
  \nonumber\\
  &  & \left. + \left( \mathe^{- i\tmmathbf{Q} \cdot \mathsf{X} / \hbar}
  \mathsf{A} \mathe^{i\tmmathbf{Q} \cdot \mathsf{X} / \hbar} -
  \mathsf{\mathsf{A}} \right) L_j \left( \mathsf{P} ; \tmmathbf{Q} \right)
  \omega^{\star}_j \left( \tmmathbf{Q} \right) \right] \nonumber\\
  &  &  \nonumber\\
  &  & + \int \mathd \mu \left( \tmmathbf{Q} \right) | \omega_j \left(
  \tmmathbf{Q} \right) |^2 \sum_j \left[ \mathe^{- i\tmmathbf{Q} \cdot
  \mathsf{X} / \hbar} \mathsf{A} \mathe^{i\tmmathbf{Q} \cdot \mathsf{X} /
  \hbar} - \mathsf{A} - \frac{i}{\hbar} \frac{\left[ \mathsf{A}, \tmmathbf{Q}
  \cdot \mathsf{X}] \right.}{1 + Q^2 / Q^2_0} \right] . \nonumber
\end{eqnarray}
Such expressions can cover a huge variety of physical situations, accounting
for both dissipative and decoherence effects. Some rough insight can be gained
considering the dummy integration label $\tmmathbf{Q}$ as a momentum. The
dynamics of the open system, in our case the centre of mass of a tracer
particle, is thus described by an interaction only characterized by the
momentum transfers between system and environment, taking place e.g. as a
consequence of collisions, thus complying with translational invariance. The
unitary operators $\exp \left( i\tmmathbf{Q} \cdot \mathsf{X} / \hbar \right)$
appearing in the Poisson part describe in fact a momentum kick, with rates
which are not only given by functions of the momentum transfer $\tmmathbf{Q}$
itself, but also depend on the momentum operator $\mathsf{P}$, thus becoming
dynamic quantities. \ This is in particular necessary in order to correctly
describe phenomena like energy transfer and approach to equilibrium. The
Gaussian part corresponds to a dynamics arising as a consequence of a big
number of small momentum transfers, leading to a diffusive behavior.

An interesting limiting situation appears if we neglect dissipative effects
and therefore the dynamics of the momentum operator, apart from its appearance
in the free kinetic term, so that, also switching to the preadjoint mapping in
Schr\"odinger picture, and assuming $\mu \left( \tmmathbf{Q} \right)$ to be
absolutely continuous with respect to the Lebesgue measure, the two
contributions can be written
\begin{eqnarray}
  \mathcal{L}_G [\rho] & = & - \frac{i}{\hbar}  \sum_{i = 1}^3 \mathbf{b}_i [
  \mathsf{X}_i, \rho] - \sum_{i, j = 1}^3 \frac{1}{2} \tmmathbf{D}_{\tmop{ij}}
  [ \mathsf{X}_i, [ \mathsf{X}_j, \rho]]  \label{eq:decoh1}\\
  \mathcal{L}_P [\rho] & = & \int d\tmmathbf{Q}| \lambda (\tmmathbf{Q}) |^2 
  \left[ e^{i\tmmathbf{Q} \cdot \mathsf{X} / \hbar} \rho e^{- i\tmmathbf{Q}
  \cdot \mathsf{X} / \hbar} - \rho - \frac{i}{\hbar}  \frac{[\tmmathbf{Q}
  \cdot \mathsf{X}, \rho]}{1 + Q^2 / Q^2_0} \right]  \label{eq:decoh2}
\end{eqnarray}
where $\mathbf{b}_{} \in \mathbb{R}, \tmmathbf{D}_{} \geq 0$, \ and the
integration measure satisfies the L\'evy condition
\begin{eqnarray}
  \int d\tmmathbf{Q}| \lambda (\tmmathbf{Q}) |^2 \frac{Q^2}{1 + Q^2} < \infty
  . &  &  \label{eq:levi}
\end{eqnarray}
It is very convenient to write the contributions given by Eq.
(\ref{eq:decoh1}) and Eq. (\ref{eq:decoh2}) in the position representation,
leading to the simple expression
\begin{eqnarray}
  \langle \tmmathbf{X}| \mathcal{L}_G [\rho] + \mathcal{L}_P [\rho]
  |\tmmathbf{Y} \rangle & = & - \Psi \left( \tmmathbf{X}-\tmmathbf{Y} \right)
  \langle \tmmathbf{X}| \rho |\tmmathbf{Y} \rangle,  \label{eq:xrepr}
\end{eqnarray}
where according to Eq. (\ref{eq:decoh1}) and Eq. (\ref{eq:decoh2}) we have
introduced the function
\begin{eqnarray}
  \Psi \left( \tmmathbf{X}-\tmmathbf{Y} \right) & = & \frac{i}{\hbar}
  \text{$\mathbf{b}_{}$} \cdot \left( \tmmathbf{X}-\tmmathbf{Y} \right) +
  \frac{1}{2} \left( \tmmathbf{X}-\tmmathbf{Y} \right)^T \cdot \tmmathbf{D}
  \cdot \left( \tmmathbf{X}-\tmmathbf{Y} \right)  \label{eq:psi}\\
  &  & - \int d\tmmathbf{Q}| \lambda (\tmmathbf{Q}) |^2  \left[
  e^{i\tmmathbf{Q} \cdot \left( \tmmathbf{X}-\tmmathbf{Y} \right) / \hbar} - 1
  - \frac{i}{\hbar}  \frac{\tmmathbf{Q} \cdot \left( \tmmathbf{X}-\tmmathbf{Y}
  \right)}{1 + Q^2 / Q^2_0} \right], \nonumber
\end{eqnarray}
only depending on the difference $\tmmathbf{X}-\tmmathbf{Y}$ due to
translational invariance. The action of the contributions given by Eq.
(\ref{eq:decoh1}) and Eq. (\ref{eq:decoh2}) in the position representation is
therefore very simple, it only amounts to multiplying the matrix elements of
the statistical operator by a function of the particular form (\ref{eq:psi}),
whose general properties as we shall see naturally account for a description
of decoherence.

\section{Decoherence and L\'evy processes\label{sec:lp}\label{sec:lp}}

The master equation corresponding to Eq. (\ref{eq:decoh1}) and Eq.
(\ref{eq:decoh2}) can be easily solved in the position representation, giving
a dynamics which only changes the initial statistical operator by a
multiplicative time dependent factor
\begin{eqnarray}
  \langle \tmmathbf{X}| \rho_t |\tmmathbf{Y} \rangle & = & \mathe^{- t \Psi
  \left( \tmmathbf{X}-\tmmathbf{Y} \right)} \langle \tmmathbf{X}| \rho_0
  |\tmmathbf{Y} \rangle .  \label{eq:sol}
\end{eqnarray}
A key point is now the observation that Eq. (\ref{eq:psi}) actually gives the
general expression of the characteristic exponent appearing in the
characteristic function of a L\'evy process, corresponding to the celebrated
L\'evy-Khintchine formula {\cite{Feller1971}}. As a consequence the function
\begin{eqnarray}
  \Phi \left( t, \tmmathbf{X}-\tmmathbf{Y} \right) & = & \mathe^{- t \Psi
  \left( \tmmathbf{X}-\tmmathbf{Y} \right)}  \label{eq:cf}
\end{eqnarray}
gives the general possible expressions for the characteristic function of a
classical L\'evy process, different processes, e.g. Gaussian, Poisson,
compound Poisson or L\'evy stable processes arising corresponding to the
different possible values of $\mathbf{b}, \tmmathbf{D}_{}$ and of the positive
weight $| \lambda (\tmmathbf{Q}) |^2$ in the measure. These different L\'evy
processes intuitively correspond to the different ways according to which
momentum is transferred to the test particle as a consequence of interaction
with the environment. Thus for example a Poisson process corresponds to a
situation in which the different possible interaction events are characterized
by a fixed momentum transfer, given by the height of the jumps in the Poisson
process. More generally a physically realistic situation involves a compound
Poisson process, characterized by the fact that the momentum transfer in the
single interaction events is not a deterministic quantity, but it is itself
described by a probability density, depending on the detail of the
microscopic interaction mechanism, according to which the Poisson process is
composed.

The function $\Phi \left( t, \tmmathbf{X}-\tmmathbf{Y} \right)$ is a
characteristic function, so that it has the following interesting properties,
explaining why Eq. (\ref{eq:sol}) generally gives a well defined master
equation describing loss of coherence in the position representation:
\begin{itemizedot}
  \item $\Phi \left( t, 0 \right) = 1$
  
  \item $\left| \Phi \left( t, \tmmathbf{X}-\tmmathbf{Y} \right) \right|
  \leqslant 1$
  
  \item $\Phi \left( t, \tmmathbf{X}-\tmmathbf{Y} \right)$ is positive
  definite
  
  \item $\Phi \left( t, \tmmathbf{X}-\tmmathbf{Y} \right) \longrightarrow 0$
  for $t \rightarrow \infty$
  
  \item $\Phi \left( t, \tmmathbf{X}-\tmmathbf{Y} \right) \longrightarrow 0$
  for $\left( \tmmathbf{X}-\tmmathbf{Y} \right) \rightarrow \infty$, provided
  there exists a probability density.
\end{itemizedot}
These properties typical of characteristic functions {\cite{Lukacs1966a}}
automatically entail that the diagonal matrix elements in the position
representation are not affected with elapsing time, thus preserving
normalization of the statistical operator, while the off-diagonal matrix
elements are generally suppressed as expected due to decoherence. Furthermore
for a fixed spatial distance $\tmmathbf{X}-\tmmathbf{Y}$ the off-diagonal
matrix elements in the position representation are fully suppressed for long
enough interaction times, while for a fixed interaction time these
off-diagonal matrix elements only go to zero if the associated process admits
a proper probability density, which is not the case e.g. for a compound
Poisson process. Depending on the particular process describing the random
momentum transfers in each scattering event different characteristic functions
appear, corresponding to different behaviors in the suppression of the
off-diagonal matrix elements for large spatial separations. The function
$\left| \Phi \left( t, \tmmathbf{X}-\tmmathbf{Y} \right) \right|$, which is
responsible for the loss of visibility in interferometric experiments testing
decoherence, for a fixed interaction time $t$ might monotonically decrease to
zero for growing values of $\tmmathbf{X}-\tmmathbf{Y}$, or also oscillate and
reach asymptotically a finite value corresponding to a residual coherence.
These quite different behaviors, corresponding to a more or less effective
decoherence effect, are all encoded in the possible expressions of the
characteristic function $\Phi$. Application of this formalism to actually
realized experiments has been considered in {\cite{Vacchini2005a}}. Typical
experiments testing decoherence in a quantitative way involve an
interferometer for massive particles (such as fullerenes
{\cite{Hornberger2003a,Hackermuller2004a}} or atoms
{\cite{Kokorowski2001a,Uys2005a}}), in which the interfering particle is
exposed to some environment during the time of flight, such as a background
gas, a laser field or even the internal degrees of freedom of the interfering
particle itself.

\section{Decoherence in momentum and position for a massive tracer
particle\label{sec:mp}}

The general structure of translation-covariant quantum dynamical semigroups
allows for the description of decoherence effects provided one considers the
behavior in time of the so called coherences, that is to say the off-diagonal
matrix elements of the statistical operator in a given basis, selected by the
dynamics itself or by the observation which can be performed on the open
system. For the considered massive particle interacting with some environment
the natural basis are given by momentum or position. In order to describe both
phenomena we obviously cannot neglect the momentum dynamics as implicitly done
going over from Eq. (\ref{eq:g+p}) to Eq. (\ref{eq:decoh1}) and Eq.
(\ref{eq:decoh2}). We therefore need a physical example of realization of the
general structure Eq. (\ref{eq:g+p}), as given by the quantum version of the
classical linear Boltzmann equation
{\cite{Vacchini2000a,Vacchini2001b,Vacchini2002a,Hornberger2006b,Hornberger2007c}}.
Such a master equation describes the dynamics of a quantum test particle
interacting through collisions with a homogeneous gas, thus providing a
quantum counterpart of the classical linear Boltzmann equation. For the case
of a scattering cross section $\sigma \left( \tmmathbf{Q} \right)$ only
depending on the momentum transfer the equation can be written
\begin{eqnarray}
  \mathcal{L} \left[ \rho \right] & = & \frac{n_{\tmop{gas}} }{m^2_{\ast}}
  \bigintlim \mathd \tmmathbf{Q} \sigma \left( \tmmathbf{Q} \right) \left[
  \mathe^{i\tmmathbf{Q} \cdot \mathsf{X} / \hbar} \sqrt{S \left( \tmmathbf{Q},
  \mathsf{P} \right)} \rho \sqrt{S \left( \tmmathbf{Q}, \mathsf{P} \right)}
  \mathe^{- i\tmmathbf{Q} \cdot \mathsf{X} / \hbar}  \label{eq:qdsf} \right.\\
  &  & \left. - \frac{1}{2} \left\{ S \left( \tmmathbf{Q}, \mathsf{P}
  \right), \rho \right\} \right], \nonumber
\end{eqnarray}
with $n_{\tmop{gas}}$ the density of gas particles with mass $m$, $M$ the mass
of the test particle, $m_{\ast} = mM / \left( m + M \right)$ the reduced mass,
$S \left( \tmmathbf{Q}, \mathsf{P} \right)$ a two-point correlation function
of the gas known as dynamic structure factor and explicitly given by
\begin{eqnarray}
  S_{} \left( \mathbf{\tmmathbf{Q}}, \tmmathbf{P} \right) & = &
  \sqrt{\frac{\beta m}{2 \pi}} \frac{1}{Q} \exp \left( - \frac{\beta}{8 m}
  \frac{(Q^2 + 2 mE \left( \tmmathbf{Q}, \tmmathbf{P} \right))^2}{Q^2}
  \right),  \label{eq:sqp}
\end{eqnarray}
with
\begin{eqnarray}
  &  & E \left( \tmmathbf{Q}, \tmmathbf{P} \right) = \frac{\left(
  \tmmathbf{P}+\tmmathbf{Q} \right)^2}{2 M} - \frac{P^2}{2 M} = \frac{Q^2}{2
  M} + \frac{\tmmathbf{Q} \cdot \tmmathbf{P}}{M}  \label{eq:etransfer}
\end{eqnarray}
the energy transfer in the single collision and $\beta = 1 / \left(
k_{\text{B}} T \right)$. We are not going to delve on details of the structure
of such an equation. We only point out that it actually provides an example of
translation-covariant master equation complying with the general mathematical
result. We are however interested to show that such a structure actually
describes decoherence phenomena in both momentum and position. In fact while
the classical linear Boltzmann equation only describes dissipative effects,
corresponding to the behavior of populations in momentum space, that is the
diagonal matrix elements in the momentum representation of Eq.
(\ref{eq:qdsf}), the quantum master equation also describes coherences and
therefore possibly interference phenomena and suppression thereof as a
consequence of the dynamics, provided suitable quantum states given by linear
superpositions states are considered.

Looking at coherence in momentum space implies considering coherent
superpositions of momentum eigenstates. Such highly non classical motional
states can show interference effects which are expected to be suppressed as a
consequence of the interaction with the environment. As a
consequence matrix elements of the form $\langle \tmmathbf{P}| \rho
|\tmmathbf{P}' \rangle$ are quickly suppressed for $\tmmathbf{P} \neq
\tmmathbf{P}'$, so that for long enough times the dynamics only affects the
behavior of the probability density $\langle \tmmathbf{P}| \rho |\tmmathbf{P}
\rangle$, and the master equation Eq. (\ref{eq:qdsf}) goes effectively over to
a classical rate equation for such a probability density. Due to the
complexity of Eq. (\ref{eq:qdsf}) obtaining an analytical solution is hardly
feasible, so that the natural strategy is to numerically solve the master
equation, relying on a so called unraveling of the master equation itself
{\cite{Breuer2007}}, to be solved by means of Monte Carlo methods. In this
case setting
\begin{eqnarray}
  V \left( \tmmathbf{Q} \right) & = & \mathe^{i\tmmathbf{Q} \cdot \mathsf{X} /
  \hbar} \sqrt{\frac{n_{\tmop{gas}} }{m^2_{\ast}} \sigma \left( \tmmathbf{Q}
  \right) S_{} \left( \tmmathbf{Q}, \mathsf{P} \right)},  \label{eq:b}
\end{eqnarray}
one can consider the following stochastic differential equation for the
stochastic wave vector $\psi \left( t \right)$
\begin{eqnarray}
  \mathd | \psi \left( t \right) \rangle & = & \left[ - \frac{1}{2} \bigintlim
  \mathd \tmmathbf{Q}V^{^{\dag}} \left( \tmmathbf{Q} \right) V \left(
  \tmmathbf{Q} \right) + \frac{1}{2} \bigintlim \mathd \tmmathbf{Q}\|V \left(
  \tmmathbf{Q} \right) | \psi \left( t \right) \rangle \|^2 \right] | \psi
  \left( t \right) \rangle \tmop{dt} \nonumber\\
  &  & + \bigintlim \mathd \tmmathbf{Q} \left[ \frac{V \left( \tmmathbf{Q}
  \right) | \psi \left( t \right) \rangle}{\|V \left( \tmmathbf{Q} \right) |
  \psi \left( t \right) \rangle \|} - | \psi \left( t \right) \rangle \right]
  \tmop{dN}_{\tmmathbf{Q}} \left( t \right),  \label{eq:d}
\end{eqnarray}
where the field of increments satisfies
\begin{eqnarray}
  \tmop{dN}_{\tmmathbf{Q}} \left( t \right) \tmop{dN}_{\tmmathbf{Q}'} \left( t
  \right) & = & \delta^3 \left( \tmmathbf{Q}-\tmmathbf{Q}' \right)
  \tmop{dN}_{\tmmathbf{Q}} \left( t \right) \nonumber\\
  \mathbbm{E} \left[ \tmop{dN}_{\tmmathbf{Q}} \left( t \right) \right] & = &
  \|V \left( \tmmathbf{Q} \right) | \psi \left( t \right) \rangle \|^2
  \tmop{dt}, \nonumber
\end{eqnarray}
so that indeed the solutions of the stochastic differential equation
(\ref{eq:d}) provide unravelings of the master equation Eq. (\ref{eq:qdsf}),
in the sense that
\begin{eqnarray}
  \rho \left( t \right) & = & \mathbbm{E} \left[ | \psi \left( t \right)
  \rangle \langle \psi \left( t \right) | \right] . \nonumber
\end{eqnarray}
Despite the formal complexity of Eq. (\ref{eq:d}), for initial states given by
momentum eigenvectors one can develop a simple algorithm to study the dynamics
of such states, essentially corresponding to the Gillespie algorithm
{\cite{Gillespie1992}}, leading to a pure jump process in momentum space. On
similar grounds one can also study the dynamics of coherent superpositions of
the form
\begin{eqnarray}
  | \psi \left( 0 \right) \rangle & = & \alpha_1 \left( 0 \right)
  |\tmmathbf{P}_1 \rangle + \alpha_2 \left( 0 \right) |\tmmathbf{P}_2 \rangle,
  \nonumber
\end{eqnarray}
with $\sum_{i = 1}^2 \bignone | \alpha_i \left( 0 \right) |^2 = 1$, which
evolve in time according to
\begin{eqnarray}
  | \psi \left( t \right) \rangle & = & \alpha_1 \left( t \right)
  |\tmmathbf{P}_1 \left( t \right) \rangle + \alpha_2 \left( t \right)
  |\tmmathbf{P}_2 \left( t \right) \rangle, \nonumber
\end{eqnarray}
where again $\sum_{i = 1}^2 \bignone | \alpha_i \left( t \right) |^2 = 1$. An
estimate of loss of coherence can be obtained studying the quantity
\begin{eqnarray}
  C \left( t \right) & = & \mathbbm{E} \left[ \frac{| \alpha_1 \left( t
  \right) \alpha_2^{\star} \left( t \right) |}{| \alpha_1 \left( 0 \right)
  \alpha_2^{\star} \left( 0 \right) |} \right] . \nonumber
\end{eqnarray}
As it turns out this measure for the coherence of the state in the momentum
basis behaves for a constant scattering cross section approximately as
{\cite{Breuer2007c}}
\begin{eqnarray}
  C \left( t \right) & = & \exp \left[ - \gamma \left( |\tmmathbf{P}_1
  -\tmmathbf{P}_2 | \right) t \right],  \label{eq:lossp}
\end{eqnarray}
where the argument of the exponential is given by
\begin{eqnarray}
  \gamma \left( P \right) & = & \Lambda \left( P \right) - \Lambda_0
  \frac{\tmop{erf} \left( P \right)}{P}, \nonumber
\end{eqnarray}
with
\begin{eqnarray}
  \Lambda \left( P \right) & = & \frac{n_{\tmop{gas}} }{m^2_{\ast}} \bigintlim
  \mathd \tmmathbf{Q} \sigma S \left( \tmmathbf{Q}, \mathsf{\tmmathbf{P}}
  \right),  \label{eq:rate}
\end{eqnarray}
$\text{erf} (x) = 2 \pi^{- \frac{1}{2}} \int^x_0 \exp \left( - t^2 \right)
\mathd t$ denotes the error function, and $\Lambda_0$ is a reference
scattering rate given by $\Lambda_0 = n_{\tmop{gas}} v_{\tmop{mp}} 4 \pi
\sigma$, with $v_{\tmop{mp}}$ the most probable velocity for the gas
particles. Eq. (\ref{eq:lossp}) clearly predicts an exponential loss of
coherence in the momentum basis, depending on the relative distance in
momentum space of the states making up the coherent superposition.

For the study of decoherence in position space we can follow a different
strategy. Neglecting in Eq. (\ref{eq:qdsf}) the dynamics of the momentum, we
can replace the corresponding operator by a classical label $\tmmathbf{P}_0$
giving the mean value of the momentum of the incoming particle. The master
equation then reads
\begin{eqnarray}
  \mathcal{L} \left[ \rho \right] & = & \frac{n_{\tmop{gas}} }{m^2_{\ast}}
  \bigintlim \mathd \tmmathbf{Q} \sigma \left( \tmmathbf{Q} \right) S \left(
  \tmmathbf{Q}, \mathsf{\tmmathbf{P}_0} \right) \left[ \mathe^{i\tmmathbf{Q}
  \cdot \mathsf{X} / \hbar} \rho \mathe^{- i\tmmathbf{Q} \cdot \mathsf{X} /
  \hbar} - \rho \right],  \label{eq:qdsfp}
\end{eqnarray}
corresponding to a particular realization of Eq. (\ref{eq:decoh2}).
Considering a constant scattering cross section and defining the rate $\Lambda
\left( P_0 \right)$ according to (\ref{eq:rate}) one can introduce the
following characteristic function
\begin{eqnarray}
  \Phi_S \left( \tmmathbf{X} \right) & = & \frac{n_{\tmop{gas}}
  \sigma}{m^2_{\ast}  \text{$\Lambda \left( P_0 \right)$}} \bigintlim \mathd
  \tmmathbf{Q}S \left( \tmmathbf{Q}, \mathsf{\tmmathbf{P}_0} \right)
  \mathe^{i\tmmathbf{Q} \cdot \tmmathbf{X}/ \hbar}, \nonumber
\end{eqnarray}
so that the master equation (\ref{eq:qdsfp}) can be solved in the position
representation as in (\ref{eq:sol}), leading to
\begin{eqnarray}
  \langle \tmmathbf{X}| \rho_t |\tmmathbf{Y} \rangle & = & \exp \left( -
  \Lambda_0 \frac{2}{\sqrt{\pi}}  \left[ 1 - \Phi_S \left(
  \tmmathbf{X}-\tmmathbf{Y} \right) \right] t \right) \langle \tmmathbf{X}|
  \rho_0 |\tmmathbf{Y} \rangle,  \label{eq:aa}
\end{eqnarray}
where according to the general framework presented in Sect. \ref{sec:lp} the
characteristic function of a compound Poisson process appears. A suitable
measure of decoherence is given in this case by
\begin{eqnarray}
  D \left( t \right) & = & \frac{\langle \tmmathbf{X}| \rho_t |\tmmathbf{Y}
  \rangle}{\langle \tmmathbf{X}| \rho_0 |\tmmathbf{Y} \rangle} . \nonumber
\end{eqnarray}
For a test particle slower than the gas particles, so that $P_0 \ll
Mv_{\tmop{mp}}$, one has
\begin{eqnarray}
  \Phi_S \left( \tmmathbf{X} \right) & \approx &  \text{}_1 F_1 \left( 1,
  \frac{3}{2} ; - 4 \pi \frac{X^2}{\lambda^2_{\tmop{th}}} \right), \nonumber
\end{eqnarray}
with $\lambda^{}_{\tmop{th}}$ the thermal de Broglie wavelength of the gas
particles given by \ $\lambda^{}_{\tmop{th}} = \sqrt{2 \pi \beta \hbar^2 /
m}$, and $\text{}_1 F_1$ the confluent hypergeometric function, so that
\begin{eqnarray}
  D \left( t \right) & = & \exp \left( - \Lambda_0 \frac{2}{\sqrt{\pi}} 
  \left[ 1 - \text{}_1 F_1 \left( 1, \frac{3}{2} ; - 4 \pi
  \frac{X^2}{\lambda^2_{\tmop{th}}} \right) \right] t \right), \nonumber
\end{eqnarray}
which for spatial distances above the thermal de Broglie wavelength $X \gg
\lambda^{}_{\tmop{th}}$ is well approximated by a fixed decoherence rate $D
\left( t \right) =$exp$\left( - 2 \Lambda_0 t / \sqrt{\pi} \right)$,
expressing the fact that for large enough distances off-diagonal matrix
elements in the position representation are uniformly suppressed.

\section{Conclusions and outlook\label{sec:co}}

We have given a brief presentation of how quantum dynamical semigroups can be
useful for the description of decoherence in quantum mechanics, as also
pursued in {\cite{Rebolledo2005a,Rebolledo2005b}}, coping in a quantitative
way with experimentally realizable situations. This has been obtained relying
on a characterization of translation-covariant quantum dynamical semigroups,
leading to a quantum non-commutative generalization of the L\'evy-Khintchine
formula. When applied to the study of decoherence, neglecting dissipative
phenomena, such a structure leads to a description of loss of coherence with a
wide variety of possible behaviors, each corresponding to the characteristic
function of a classical L\'evy process. Despite pursued within the framework
of the Markov assumption, thus supposing that the dynamics does not entail
memory effects, the approach to the description of decoherence building on
covariance properties, recently also followed in {\cite{Clark2008a}}, can be
of more general validity, as it appears from recent results pointing to a
generalization of the Lindblad structure for the description of a class of
non-Markovian evolutions {\cite{Breuer2007a}}.

\section{Acknowledgments}

The author is grateful to the organizers for the kind invitation and
hospitality at CIMAT. The work was partially supported by the Italian MUR
under PRIN2005.


\begin{thebibliography}{10}
  \bibitem[1]{Streater2000a}R.~F. Streater, \tmtextit{J. Math. Phys.}
  \tmtextbf{41}, 3556 (2000).
  
  \bibitem[2]{Holevo2001a}A.~S. Holevo, \tmtextit{Statistical Structure of
  Quantum Theory} (Springer, Berlin, 2001).
  
  \bibitem[3]{Alicki2001a}R.~Alicki and M.~Fannes, \tmtextit{Quantum dynamical
  systems} (Oxford University Press, Oxford, 2001).
  
  \bibitem[4]{Holevo1993a}A.~S. Holevo, \tmtextit{Rep. Math. Phys.}
  \tmtextbf{32}, 211 (1993).
  
  \bibitem[5]{Holevo1993b}A.~S. Holevo, \tmtextit{Rep. Math. Phys.}
  \tmtextbf{33}, 95 (1993).
  
  \bibitem[6]{Holevo1995a}A.~S. Holevo, \tmtextit{Izv. Math.} \tmtextbf{59},
  427 (1995).
  
  \bibitem[7]{Holevo1996a}A.~S. Holevo, \tmtextit{J.~Math. Phys.}
  \tmtextbf{37}, 1812 (1996).
  
  \bibitem[8]{Vacchini2001b}B.~Vacchini, \tmtextit{J.~Math. Phys.}
  \tmtextbf{42}, 4291 (2001).
  
  \bibitem[9]{Vacchini2002b}B.~Vacchini, \tmtextit{J.~Math. Phys.}
  \tmtextbf{43}, 5446 (2002).
  
  \bibitem[10]{Vacchini2005a}B.~Vacchini, \tmtextit{Phys. Rev. Lett.}
  \tmtextbf{95}, p. 230402 (2005).
  
  \bibitem[11]{Joos2003a}E.~Joos, H.~D. Zeh, C.~Kiefer, D.~Giulini, J.~Kupsch
  and I.-O. Stamatescu, \tmtextit{Decoherence and the Appearance of a
  Classical World in Quantum Theory}, 2nd edn. (Springer, Berlin, 2003).
  
  \bibitem[12]{Schlosshauer2007a}M.~Schlosshauer, \tmtextit{Decoherence and
  the Quantum-To-Classical Transition} (Springer-Verlag, Berlin, 2007).
  
  \bibitem[13]{Alicki1987a}R.~Alicki and K.~Lendi, \tmtextit{Quantum Dynamical
  Semigroups and Applications} (Springer, Berlin, 1987).
  
  \bibitem[14]{Alicki2002b}R.~Alicki, Invitation to quantum dynamical
  semigroups, in \tmtextit{Dynamical semigroups: Dissipation, chaos, quanta},
  eds. P.~Garbaczewski and R.~Olkiewicz, Lecture Notes in Physics, Vol.~597
  (Springer-Verlag, Berlin, 2002) pp. 239--264.
  
  \bibitem[15]{Breuer2007}H.-P. Breuer and F.~Petruccione, \tmtextit{The
  Theory of Open Quantum Systems} (Oxford University Press, Oxford, 2007).
  
  \bibitem[16]{Gorini1976a}V.~Gorini, A.~Kossakowski and E.~C.~G. Sudarshan,
  \tmtextit{J.~Math. Phys.} \tmtextbf{17}, 821 (1976).
  
  \bibitem[17]{Lindblad1976a}G.~Lindblad, \tmtextit{Comm. Math. Phys.}
  \tmtextbf{48}, 119 (1976).
  
  \bibitem[18]{Vacchini2007c}B.~Vacchini, to appear in Lecture Notes
    in Physics [arXiv:quant-ph/0707.0603].
  
  \bibitem[19]{Holevo1995b}A.~S. Holevo, \tmtextit{J. Funct. Anal.}
  \tmtextbf{131}, 255 (1995).
  
  \bibitem[20]{Feller1971}W.~Feller, \tmtextit{An introduction to probability
  theory and its applications. Vol. II} (John Wiley \& Sons Inc., New York,
  1971).
  
  \bibitem[21]{Lukacs1966a}L.~Lukacs, \tmtextit{Characteristic Functions}
  (Griffin, London, 1966).
  
  \bibitem[22]{Hornberger2003a}K.~Hornberger, S.~Uttenthaler, B.~Brezger,
  L.~Hackerm\"uller, M.~Arndt and A.~Zeilinger, \tmtextit{Phys. Rev. Lett.}
  \tmtextbf{90}, p. 160401 (2003).
  
  \bibitem[23]{Hackermuller2004a}L.~Hackerm\"uller, K.~Hornberger, B.~Brezger,
  A.~Zeilinger and M.~Arndt, \tmtextit{Nature} \tmtextbf{427}, 711 (2004).
  
  \bibitem[24]{Kokorowski2001a}D.~A. Kokorowski, A.~D. Cronin, T.~D. Roberts
  and D.~E. Pritchard, \tmtextit{Phys. Rev. Lett.} \tmtextbf{86}, p. 2191
  (2001).
  
  \bibitem[25]{Uys2005a}H.~Uys, J.~D. Perreault and A.~D. Cronin,
  \tmtextit{Phys. Rev. Lett.} \tmtextbf{95}, p. 150403 (2005).
  
  \bibitem[26]{Vacchini2000a}B.~Vacchini, \tmtextit{Phys. Rev. Lett.}
  \tmtextbf{84}, 1374 (2000).
  
  \bibitem[27]{Vacchini2002a}B.~Vacchini, \tmtextit{Phys. Rev.~E}
  \tmtextbf{66}, p. 027107 (2002).
  
  \bibitem[28]{Hornberger2006b}K.~Hornberger, \tmtextit{Phys. Rev. Lett.}
  \tmtextbf{97}, p. 060601 (2006).
  
  \bibitem[29]{Hornberger2007c}K.~Hornberger and B.~Vacchini, \tmtextit{Phys. Rev.~A}
  \tmtextbf{77}, p. 022112 (2008).
  
  \bibitem[30]{Gillespie1992}D.~T. Gillespie, \tmtextit{Markov Processes}
  (Academic Press, Boston, 1992).
  
  \bibitem[31]{Breuer2007c}H.-P. Breuer and B.~Vacchini, \tmtextit{Phys.
  Rev.~E} \tmtextbf{76}, p. 036706 (2007).
  
  \bibitem[32]{Rebolledo2005a}R.~Rebolledo, \tmtextit{Ann. Inst. H. Poincar\'e
  Probab. Statist.} \tmtextbf{41}, 349 (2005).
  
  \bibitem[33]{Rebolledo2005b}R.~Rebolledo, \tmtextit{Open Syst. Inf. Dyn.}
  \tmtextbf{12}, 37 (2005).
  
  \bibitem[34]{Clark2008a}J.~Clark, \tmtextit{J. Math. Phys.} \tmtextbf{49}, 052103 (2008).
  
  \bibitem[35]{Breuer2007a}H.-P. Breuer, \tmtextit{Phys. Rev.~A}
  \tmtextbf{75}, p. 022103 (2007).
\end{thebibliography}
\end{document}